\documentclass[12pt]{iopart}

\usepackage{iopams}

\usepackage{graphicx}
\usepackage{theorem}

\theoremstyle{plain}
\newtheorem{Thm}{Theorem}
\newtheorem{Lem}[Thm]{Lemma}
\newtheorem{Cor}[Thm]{Corollary}
\newtheorem{Prop}[Thm]{Proposition}

\newtheorem{Def}{Definition}

\theorembodyfont{\rmfamily}
\newtheorem{Proof}{Proof}

\begin{document}

\title{Vertex operator for the ultradiscrete KdV equation}

\author{Yoichi Nakata}
\address{Graduate School of Mathematical Sciences, The University of Tokyo, 3-8-1 Komaba, Meguro-ku, 153-8914 Tokyo, Japan}
\ead{ynakata@ms.u-tokyo.ac.jp}
\begin{abstract}
We propose an ultradiscrete analogue of the vertex operator in the case of the ultradiscrete KdV equation, which maps $N$-soliton solutions to $N+1$-soliton ones. 
\end{abstract}

\vspace{2pc}

\bigskip

The ultradiscrete KdV equation (udKdVeq)
\begin{equation}\label{bilinear}
	F^t_j + F^{t+2}_{j+1} = \max ( F^{t+2}_j + F^t_{j+1} -2R, F^{t+1}_j + F^{t+1}_{j+1} ) \qquad (R \ge 0),
\end{equation}
was first proposed in \cite{TTMS} to describe the dynamics of the Box and Ball System (BBS) \cite{TS}, which is a cellular automaton with soliton-like behaviour. It has a rich structure, including the existence of $N$-soliton solutions and an infinite amount of conserved quantities \cite{NTS} like most ordinary soliton equations.

A procedure to obtain a new solution of a soliton equation from a given one, is known under the name of  B\"acklund transformation (BT) and is generally expressed in the form of the differential equations. The ultradiscrete analogue of the BT for the KdV equation in the case of the udKdVeq is presented in \cite{IKMS}.

The vertex operator is an operator representation of a BT and was first presented in \cite{DJKM} to describe the structure of the solutions and the symmetries of soliton systems. It maps $N$-soliton solutions to $N+1$-soliton ones and all soliton solutions can in fact be generated by repeated application for the vertex operator.

In this letter, we propose an ultradiscrete analogue of vertex operator in the case of the ultradiscrete KdV equation and prove that the functions generated by this operator indeed solve the udKdVeq.

\bigskip
\bigskip

First, we define the vertex operator and consider the functions generated by this operator.
\begin{Def}
The multi-variable function $F$ and the vertex operator $X$ are defined as follows:
\begin{enumerate}
\item The null variable function $F(;)$ is defined as:
\begin{equation}
	F(;)=0.
\end{equation}
\item The vertex operator $X$ depends one real parameter $\Omega_1\ge 0$ and a function $\eta_1(\Omega_1)$ of $\Omega_1$, and maps the null variable function $F(;)$ to
\begin{equation}
	X(\Omega_1, \eta_1)F(;) := \max( -\eta_1, \eta_1 ).
\end{equation}
Accordingly, the one variable function $F(\Omega_1; \eta_1)$ is defined as:
\begin{equation}
	F(\Omega_1; \eta_1) := X(\Omega_1, \eta_1)F(;).
\end{equation}
\item For general $N\ge1$, the vertex operator $X$ maps an $N$ variable function $F(\Omega_1, \ldots, \Omega_{N}; \eta_{1}, \ldots, \eta_{N})$
(written as $F(\boldsymbol{\Omega}; \boldsymbol{\eta})$ for brevity) to the $N+1$ variable function as:
\begin{eqnarray}\label{symexformula}
\fl	X(\Omega_{N+1}, \eta_{N+1})F(\boldsymbol{\Omega}; \boldsymbol{\eta}) &:= \max\big(-\eta_{N+1}+F(\boldsymbol{\Omega}; \boldsymbol{\eta}+\min{}_{\Omega_{N+1}}(\mathbf{\Omega})), \nonumber\\
																	 &\qquad\qquad \eta_{N+1}+F(\boldsymbol{\Omega}; \boldsymbol{\eta}-\min{}_{\Omega_{N+1}}(\mathbf{\Omega}))\big) \nonumber\\
																	 &=: F(\Omega_{1}, \ldots, \Omega_{N}, \Omega_{N+1}; \eta_1, \ldots, \eta_{N}, \eta_{N+1}),
\end{eqnarray}
where
\begin{equation}
 \min{}_{\Omega_{N+1}}(\boldsymbol{\Omega})={}^t(\min(\Omega_1, \Omega_{N+1}), \ldots, \min(\Omega_N, \Omega_{N+1})).
\end{equation}
\end{enumerate}
\end{Def}

The parameters $\Omega$ and the functions $\eta$ in the vertex operator $X$ are in fact the amplitudes and phase parameter of the new soliton, inserted by the operator. The definition (\ref{symexformula}) indicates that all pre-existing solitons described by $F$, have their phases shifted by inserting a new soliton. 

Here we present the basic properties of the operator $X$ and the corresponding function $F$.
\begin{Prop} \label{sym}
The action of the vertex operators are commutative.
\end{Prop}

\begin{Proof}
We calculate $X(\Omega_b, \eta_b) X(\Omega_a, \eta_a) F({\boldsymbol{\Omega}}; {\boldsymbol{\eta}})$ directly following the definition (\ref{symexformula}).
\begin{eqnarray}
\fl	X(\Omega_b, \eta_b) X(\Omega_a, \eta_a) F({\boldsymbol{\Omega}}; {\boldsymbol{\eta}}) \nonumber\\
	\!\!\!\!\!\!\!\! = \max ( - \eta_a - \eta_b - \min ( \Omega_a, \Omega_b ) + F({\boldsymbol{\Omega}}; {\boldsymbol{\eta}+\min{}_{\Omega_a}({\boldsymbol{\Omega}})+\min{}_{\Omega_b}({\boldsymbol{\Omega}})}), \nonumber\\
					\eta_a - \eta_b + \min ( \Omega_a, \Omega_b ) + F({\boldsymbol{\Omega}}; {\boldsymbol{\eta}-\min{}_{\Omega_a}({\boldsymbol{\Omega}})+\min{}_{\Omega_b}({\boldsymbol{\Omega}})}), \nonumber\\
					- \eta_a + \eta_b + \min ( \Omega_a, \Omega_b ) + F({\boldsymbol{\Omega}}; {\boldsymbol{\eta}+\min{}_{\Omega_a}({\boldsymbol{\Omega}})-\min{}_{\Omega_b}({\boldsymbol{\Omega}})}), \nonumber\\
					 \eta_a + \eta_b - \min ( \Omega_a, \Omega_b ) + F({\boldsymbol{\Omega}}; {\boldsymbol{\eta}-\min{}_{\Omega_a}({\boldsymbol{\Omega}})-\min{}_{\Omega_b}({\boldsymbol{\Omega}})}))
\end{eqnarray}
From this relation it is clear that interchanging the subscripts $a$ and $b$ does not change its overall value.
\end{Proof}
Rewriting this proposition in the language of the function $F$, yields the following corollary:
\begin{Cor}\label{cor1}
The $N$ variable functions $F({\boldsymbol{\Omega}}; {\boldsymbol{\eta}})$ are invariant under the permutation of their parameters, i.e.:
\begin{equation}
	\!\!\!\!\!\!\!\!\!\!\!\!\!\!\!\!\!\! F(\Omega_{1}, \ldots, \Omega_{N}; \eta_{1}, \ldots, \eta_{N}) = F(\Omega_{\sigma(1)}, \ldots, \Omega_{\sigma(N)}; \eta_{\sigma(1)}, \ldots, \eta_{\sigma(N)}) \quad (\sigma \in \mathfrak{S}_N ).
\end{equation}
\end{Cor}

Next, let us prove that these functions are indeed solutions of the udKdVeq, by means of the recursive form (\ref{symexformula}). This property indicates that the vertex operator defined in (\ref{symexformula}) is nothing but the operator, well known from soliton theory, which maps an $N$-soliton solution to an $N+1$-soliton solution.

\begin{Thm}\label{Thm1}
The $N$ variable function $F(\boldsymbol{\Omega}; \boldsymbol{\eta})$ solves the udKdVeq (\ref{bilinear}) for 
\begin{equation}
\boldsymbol{\eta} = \mathbf{C}-j\mathbf{K}+t\boldsymbol{\Omega},
\end{equation}
where
\begin{equation}
\mathbf{K}=\min{}_{R}(\boldsymbol{\Omega})
\end{equation}
and $\mathbf{C}$ is a constant. 
\end{Thm}

\begin{Proof}
By virtue of corollary \ref{cor1}, we can fix the labels of the parameters
\begin{equation}
	\Omega_N \ge \Omega_{N-1} \ge \ldots \ge \Omega_1 \ge 0
\end{equation}
without loss of generality. By virtue of this ordering, the phase shifts in definition (\ref{symexformula}) simplify to
\begin{equation}
	\min(\Omega_{i}, \Omega_{N}) = \Omega_{i} \quad (i=1,\ldots,N-1).
\end{equation}
It should be noted that the phase shifts $\boldsymbol{\eta} \to \boldsymbol{\eta} \pm \boldsymbol{\Omega}$ are equivalent to the time shifts $t \to t\pm 1$. The $N$ variable function $F(\boldsymbol{\Omega}; \boldsymbol{\eta})$ is then reduced to $F^{(N), t}_j$, written as
\begin{eqnarray}\label{exformula}
	F^{(N), t}_j &= \max( -C_N + j K_N - t \Omega_N + F^{(N-1), t+1}_j, \nonumber\\
				 &\qquad\qquad C_N - j K_N + t \Omega_N + F^{(N-1), t-1}_j ) \qquad&( N \ge 1) \\
	F^{(0), t}_j &= 0 \qquad&( N = 0 ).
\end{eqnarray}
We shall prove the theorem inductively. It is clear that $F^{(0), t}_j$ solves the equation (\ref{bilinear}) because of the positivity of $R$. Now, let us assume that the theorem holds for $N-1$. By substituting (\ref{exformula}) in the udKdVeq (\ref{bilinear}), the left hand side can be rewritten as
\begin{eqnarray}
\fl	F^{(N), t}_{j} + F^{(N), t+2}_{j+1} = \nonumber\\
	\max \Big( - 2 C_{N} + (2j+1) K_{N} - (2t+2) \Omega_{N} + F^{(N-1), t+1}_{j}+F^{(N-1), t+3}_{j+1}, \nonumber\\
	\qquad 2 C_{N} - (2j+1) K_{N} + (2t+2) \Omega_{N} + F^{(N-1), t-1}_{j}+F^{(N-1),t+1}_{j+1}, \nonumber\\
	 \qquad  K_{N} - 2 \Omega_{N} + F^{(N-1), t-1}_{j} + F^{(N-1), t+3}_{j+1}, \nonumber\\
	 \qquad - K_{N} + 2 \Omega_{N} + F^{(N-1), t+1}_{j} + F^{(N-1), t+1}_{j+1} \Big). \label{ff1} 
\end{eqnarray}
In this expression it looks as if the maximum in (\ref{ff1}) has four arguments. However, the third argument cannot be the maximum because it is always less than the fourth one by virtue of the following lemma, in cases  $l=-1,m=2$.
\begin{Lem} \label{Lem1}
Let
\begin{equation}
	H^{(N), t}_{j} = F^{(N), t+m+2}_{j} - F^{(N), t+2}_{j+l} - F^{(N), t+m}_{j} + F^{(N), t}_{j+l},
\end{equation}
where parameters $l$ and $m$ satisfy $ l K_{N} + m \Omega_{N} \ge 0$. Then the relation
\begin{equation}
	H^{(N), t}_j \le 2( l K_{N} + m \Omega_{N})
\end{equation}
holds.
\end{Lem}

\begin{Proof}
To prove this lemma, let us introduce some fundamental properties of the $\max$ operator.
\begin{Prop}
The inequalities
\begin{eqnarray}
	\max ( a + c, b + d ) \le \max ( a, b ) + \max ( c, d ) \label{maxeq1} \\
	\max ( x, y ) - \max ( z, w ) \le \max ( x - z, y - w ) \label{maxeq2}
\end{eqnarray}
hold for arbitrary $a,b,c,d,x,y,z,w \in \mathbb{R}$
\end{Prop}

\begin{Proof}
Proof of (\ref{maxeq1}): the right hand side of the equation (\ref{maxeq1}) can be expanded to yield $\max ( a + c, a + d, b + c, b + d )$, which includes all candidates of the left hand side. 

\noindent Proof of (\ref{maxeq2}): put $a=z, b=w, c=x-z, d=y-w$ in (\ref{maxeq1}) respectively.
\end{Proof}
If we now apply
\begin{equation}
	\min ( a, b ) = - \max ( -a, -b ),
\end{equation}
we obtain the following corollary:
\begin{Cor}\label{Cor2}
The above inequalities (\ref{maxeq1}), (\ref{maxeq2}) hold even when $\max$ and $\le$ are replaced by $\min$ and $\ge$.
\end{Cor}

By employing the inequality (\ref{maxeq2}), we obtain
\begin{eqnarray}
\fl	F^{(N), t+m+2}_{j} - F^{(N), t+2}_{j+l} \le \max (-l K_{N} - m \Omega_{N} + F^{(N-1), t+k+3}_{j} - F^{(N-1), t+3}_{j+l}, \nonumber\\
				\qquad\qquad\qquad l K_{N} + m \Omega_{N} + F^{(N-1), t+k+1}_{j} - F^{(N-1), t+1}_{j+l} ) \\
\fl	F^{(N), t}_{j+l} - F^{(N), t+m}_{j} \le \max ( l K_{N} + m \Omega_{N} + F^{(N-1), t+1}_{j+l} - F^{(N-1), t+k+1}_{j}, \nonumber\\
				\qquad\qquad\qquad -l K_{N} - m \Omega_{N} + F^{(N-1), t-1}_{j+l} - F^{(N-1), t+k-1}_{j} ).
\end{eqnarray}
Adding the inequalities yields
\begin{eqnarray}
	H^{(N), t}_{j}  \le \max ( &H^{(N-1), t+1}_{j}, \nonumber\\
							   &2 ( l K_{N} + m \Omega_{N} ), \nonumber\\
							   &-2 ( l K_{N} + m \Omega_{N} ) + H^{(N-1), t+1}_{j} + H^{(N-1), t-1}_{j}, \nonumber\\
							   &H^{(N-1), t-1}_{j} ).
\end{eqnarray}
By taking into account the relation $\Omega_{N} \ge \Omega_{N-1}$, it can be shown inductively that the four arguments in this maximum are all less than $2 ( l K_{N} + m \Omega_{N} )$.
\end{Proof}

By means of the same procedure, the right hand side of equation (\ref{bilinear}) is rewritten as
\begin{eqnarray}
\fl	\max ( F^{t+2}_j + F^t_{j+1} -2R, F^{t+1}_j + F^{t+1}_{j+1} ) = \nonumber\\
	\max \Big( - 2 C_{N} - (2j+1) K_{N} - (2t+2) \Omega_{N} \nonumber\\
				\qquad + \max \big( F^{(N-1), t+3}_{j} + F^{(N-1), t+1}_{j+1} - 2R, F^{(N-1), t+2}_{j} + F^{(N-1), t+2}_{j+1}\big), \nonumber\\
				 \qquad 2 C_{N} + (2j+1) K_{N} + (2t+2) \Omega_{N} \nonumber\\
				\qquad + \max \big( F^{(N-1), t+1}_{j} + F^{(N-1), t-1}_{j+1} -2R,  F^{(N-1), t}_{j} + F^{(N-1), t+1}_{j+1}), \nonumber\\
		\qquad \max \big(K_{N} + 2 \Omega_{N} + +  F^{(N-1), t+1}_{j} + F^{(N-1), t+1}_{j+1} -2R, \nonumber\\
				\qquad \qquad K_{N} + F^{(N-1), t}_{j} + F^{(N-1), t+2}_{j+1}\big) \Big).
\end{eqnarray}
There are three arguments in the principal maximum. The first and second arguments are the same as those on  the left hand side because $F^{(N-1), t}_j$ solves the equation (\ref{bilinear}) by the assumption. It can also be shown that the last argument is also the same by employing the method which was used to prove lemma \ref{Lem1}. 

The condition which expresses the equality of the last argument to that of the left hand side then reduces:
\begin{equation}
\fl	0 = \max ( 2(K_{N} - R), 2(K_{N}-\Omega_{N}) + F^{(N-1), t}_{j} + F^{(N-1), t+2}_{j+1} - F^{(N-1), t+1}_{j} - F^{(N-1), t+1}_{j+1} )
\end{equation}
and holds by virtue of the following lemma:
\begin{Lem} \label{Lem2}
Let
\begin{equation}
	 \tilde{H}^{(N), t}_{j} = F^{(N-1), t}_{j} + F^{(N-1), t+2}_{j+1} - F^{(N-1), t+1}_{j} - F^{(N-1), t+1}_{j+1},
\end{equation}
one then has:
\begin{equation}
	 0 \le \tilde{H}^{(N), t}_{j} \le 2( \Omega_{N} - K_{N} ).
\end{equation}
\end{Lem}

\begin{Proof}
By employing the inequality (\ref{maxeq2}), we obtain
\begin{eqnarray}
\fl F^{(N-1), t}_{j} - F^{(N-1), t+1}_{j+1} \le \max (K_{N} - \Omega_{N} + F^{(N-1), t+1}_{j} - F^{(N-1), t+2}_{j+1}, \nonumber\\
					\qquad\qquad\qquad - K_{N} + \Omega_{N} + F^{(N-1), t-1}_{j} - F^{(N-1), t}_{j+l}) \\
\fl F^{(N-1), t+2}_{j+1} - F^{(N-1), t+1}_{j} \le \max - K_{N} + \Omega_{N} + F^{(N-1), t+3}_{j+1} - F^{(N-1), t+2}_{j}, \nonumber\\
					\qquad\qquad\qquad K_{N} - \Omega_{N} + F^{(N-1), t+1}_{j+1} - F^{(N-1), t}_{j+1})
\end{eqnarray}
Adding these inequalities, we obtain
\begin{eqnarray}
	\tilde{H}^{(N), t}_{j}  \le \max ( &\tilde{H}^{(N-1), t+1}_{j}, \nonumber\\
								&2 ( \Omega_{N} - K_{N} ) - \tilde{H}^{(N-1), t}_{j}, \nonumber\\
								&-2 ( \Omega_{N} - K_{N} ) + \tilde{H}^{(N-1), t-1}_{j} + \tilde{H}^{(N-1), t}_{j} + \tilde{H}^{(N-1), t+1}_{j}, \nonumber\\
								&\tilde{H}^{(N-1), t-1}_{j} ).
\end{eqnarray}
Noticing that the condition for the third argument to be the maximum of all four:
\begin{eqnarray}
\fl	K_{N} - \Omega_{N} + F^{(N-1), t+1}_{j} - F^{(N-1), t+2}_{j+1} \le - K_{N} + \Omega_{N} + F^{(N-1), t-1}_{j} - F^{(N-1), t}_{j+l} \\
\fl	- K_{N} + \Omega_{N} + F^{(N-1), t+3}_{j+1} - F^{(N-1), t+2}_{j} \ge  K_{N} - \Omega_{N} + F^{(N-1), t+1}_{j+1} - F^{(N-1), t}_{j+1},
\end{eqnarray}
reduces to
\begin{eqnarray}
	\tilde{H}^{(N-1), t+1}_{j} + \tilde{H}^{(N-1), t}_{j} \le 2 ( \Omega_{N} - K_{N} ) \\
	\tilde{H}^{(N-1), t}_{j} + \tilde{H}^{(N-1), t-1}_{j} \le 2 ( \Omega_{N} - K_{N} ),
\end{eqnarray}
it can be shown inductively that all arguments are less than $2 ( \Omega_{N} - K_{N} )$.

The positivity of $\tilde{H}^{(N), t}_j$ is proven by a similar evaluation by means of  corollary \ref{Cor2}.
\end{Proof}
\end{Proof}

Finally, we give an explicit example of the action of the vertex operator $X$, in case of the Box and Ball System. The BBS corresponds to the udKdVeq (\ref{bilinear}) by means of the dependent variable transformation
\begin{equation}
	B^t_j = \frac{1}{2}\left(F^{t+1}_j + F^t_{j+1} - F^{t+1}_{j+1} - F^t_j\right)
\end{equation}
and expressed as:
\begin{equation}
	B^{t+1}_j = \min \Big( R - B^t_j, \sum_{n=-\infty}^{j-1} (B^{t+1}_j - B^t_j) \Big).
\end{equation}
The BBS is required to satisfy the following boundary conditions: for sufficiently large j, $B^t_j = 0$. Under these conditions, it can be shown that the possible values of the variable $B^t_j$ that correspond to the $N$-soliton solutions of theorem \ref{Thm1} (i.e. constructed by means of the vertex operator $X$) are restricted to the set $\{0,1\}$.

\begin{figure}
	\includegraphics{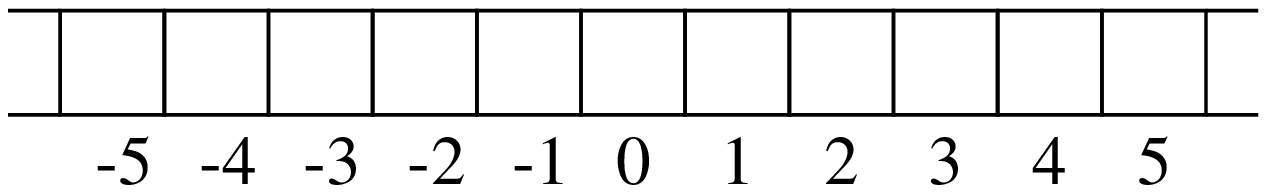} 
	\includegraphics{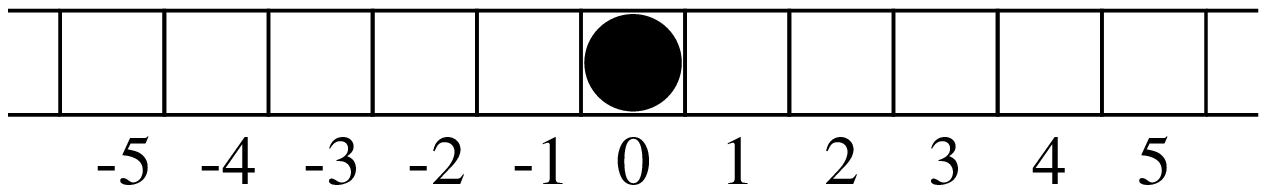} 
	\includegraphics{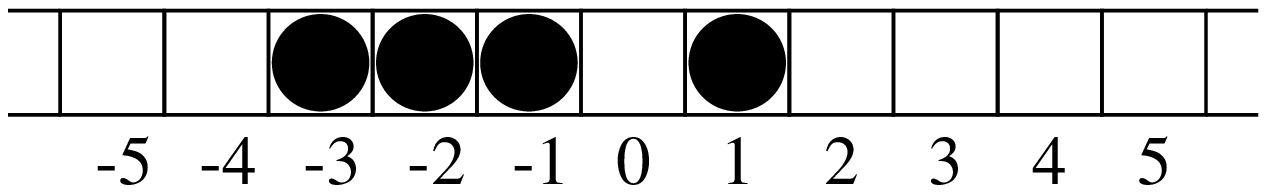}
	\includegraphics{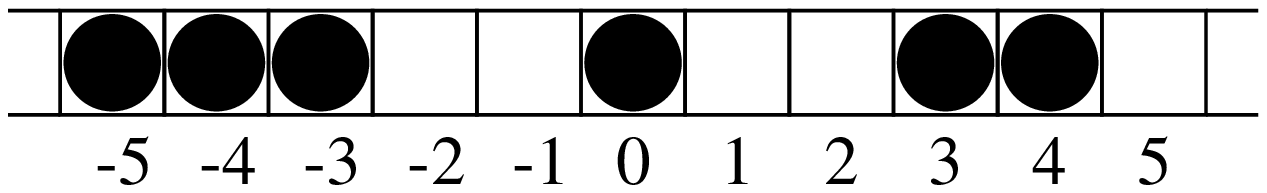}
	\caption{BBS obtained by successive application of the vertex operator.} \label{pic1}
\end{figure}

In the following figures we depict the $j$-lattice by a row of boxes, containing a ball when $B^t_j$ is equal to $1$, and empty whenever $B^t_j=0$. The lattices depicted in Figure. \ref{pic1} are the vacuum lattice, and the lattices that obtained from vacuum by the successive application of the vertex operator $X$ with parameter values $(\Omega_1, C_1)=(1,0)$, $(\Omega_2, C_2)=(3,-2)$, and $(\Omega_3, C_3)=(2,0)$.

\bigskip

In this letter we proposed an ultradiscrete analogue of the vertex operator for the ultradiscrete KdV equation and discussed its properties. We presented a recursive representation of the $N$-soliton solution generated by this operator and proved that it indeed solves the udKdVeq.

In \cite{TH}, a different technique  for describing the udKdV $N$-soliton solutions was introduced, relying on signature-free Casorati-type determinants or permanents. Although these permanents describe the same solutions (already proposed in \cite{TTMS}) as those constructed by recursive application of the ultradiscrete vertex operator, we would like to stress that the two aproaches are quite different. In fact, the vertex operator approach is closely related to the existence of certain symmetry algebras for integrable systems and the exact relation of our ultradiscrete operator to the symmetries of ultradiscrete systems is an especially interesting problem we want to address in the future. It would also be interesting to find vertex operators for other ultradiscrete soliton equations, as for example the ultradiscrete analogue of the KP equation.

\end{document}